\noindent
{\bf Holonomy invariance, orbital resonances, and kilohertz QPOs}

\vskip 1truecm

M.A. Abramowicz,      Chalmers University, Sweden;
                      and SISSA, Italy

G.J.E. Almergren,      SISSA, Italy

W. Klu{\'z}niak,      Zielona G{\'o}ra University, Poland;
                      and SISSA, Italy

A.V. Thampan,         SISSA, Italy

F. Wallinder,         {\"O}rebro University, Sweden

\vskip 1truecm

marek@fy.chalmers.se

joal@sissa.it

wlodek@camk.edu.pl

thampan@sissa.it

fredrik.wallinder@nat.oru.se

\vskip 0.3truecm
\noindent
{\bf Abstract}

\noindent
Quantized orbital structures are typical for many aspects of classical
 gravity (Newton's as well as Einstein's). 
The astronomical phenomenon of orbital resonances is a well-known example.
Recently, Rothman, Ellis and Murugan (2001) discussed quantized orbital
 structures in the novel context of a holonomy invariance of parallel transport
 in Schwarzschild geometry.
We present here yet another example of quantization of orbits,
 reflecting {\it both} orbital resonances {\it and} holonomy
 invariance.
This strong-gravity effect may already have been
 directly observed as the puzzling kilohertz quasi-periodic
 oscillations (QPOs) in the X-ray
 emission from a few accreting galactic black holes and several neutron stars.

\vskip 0.5truecm
\noindent
{\bf 1. Quantization of circular orbits: holonomy invariance}

\vskip 0.5truecm
\noindent
In a curved spacetime, consider a vector under parallel transport
 along a closed curve. The
initial and final direction of the vector are not holonomy invariant:
 in general they will
be different after one loop. But could they be holonomy invariant
 after $N$ loops? Rothman,
Ellis and Murugan [1] show that for geodesic motion along an $r =$
 const
 circle in
Schwarzshild geometry ($r$ is the standard Schwarzschild radial
 coordinate) the holonomy
invariance occurs after $N$ loops if and only if

$$ 
r = r_{\rm holonomy} = {3r_G \over 2}{1 \over {1 - n^2/m^2}},
 \eqno (1) 
$$
\noindent
where $r_G$ is the gravitational radius, $N =nm$, and 
$1 \le n < m$ are two integer numbers.
It is clear that the holonomy invariance condition leads to a
 ``quantization" of orbits.

\vskip 0.5truecm
\noindent
{\bf 2. Quantization of circular orbits: the closure condition}

\vskip 0.5truecm
\noindent
Consider now a slightly perturbed circular orbit, 
$r + \xi(t)$, where $\xi \ll r$, and $t$
is the standard Schwarzschild time coordinate. 
The equation that governs the perturbation is
that of a harmonic oscillator (see e.g. [2]),

$$
{\ddot \xi} + \omega^2_r \xi = 0,  
\eqno (2)
$$
\noindent
where a dot means $d/dt$, and $\omega_r$ is the radial epicyclic frequency.
 For stable
oscillations $\omega_r^2 > 0$, while for $\omega_r^2 < 0$
 perturbations  are clearly exponential in time and thus unstable.

\par
With Euclidean geometry of space (which is assumed in Newton's
 theory), {\it and} for a
spherically symmetric gravitational potential, orbital frequency and 
radial epicyclic frequency are always real and equal,
 $\Omega^2 =\omega_r^2 > 0$. Therefore, all slightly
perturbed circular orbits close after exactly one loop, and they
 are all stable. This
conclusion, of course, holds not only for slightly perturbed orbits 
which we are now
discussing, but for all bound, i.e., with negative energy, 
Newtonian orbits in a spherical
potential---they all are stable ellipses, as was known already by Kepler.

\par
However, $\Omega^2 \neq \omega_r^2$
if the gravitational potential is non-spherical, 
or the geometry is non-Euclidean,
which means that, in general,
 perturbed circular orbits do not
close after one loop. It may be in addition that for a range of radii
 $\omega_r^2 < 0$, which
means that orbits are unstable in this range. These facts are
well-known in the context of Schwarzschild geometry, where [2],

$$
\omega_r = \left ( 1 - {3r_G \over r} \right )^{1/2} \Omega,
\eqno (3a)
$$
$$\Omega = \left ( {r_G \over 2r}\right)^{1/2}{c \over r}.
\eqno (3b)
$$  
\noindent
and therefore 
the slightly perturbed circular orbits do not close after one loop 
(this is the reason for the relativistic advance of the perihelion
 of Mercury). In addition the
perturbations are unstable in the region
 $r < r_{ms}=3r_G$, with $r_{ms}$ being known as the
marginally stable orbit radius.

\par
It is perhaps less known, however, that the same effects do happen
 also according to Newton's
theory in a gravitational field of a sufficiently non-spherical body.
 It has been recently 
demonstrated [3] that the marginally stable orbit exists outside
 rapidly rotating Maclaurin spheroids, as well as outside strange stars 
(quark stars with an equation of state comfortably consistent
 with the main-stream views on super-dense matter) 
with a very large
quadrupole moment, but well in the Newtonian regime 
($r_{ms} >R_*>> r_{\rm G}$). Here $R_*$
is the stellar radius.

\par
Returning to the Schwarzschild geometry, let us state some obvious 
consequences of Eqs. (3). 
The slightly perturbed circular orbits close after an integral number
of loops if and only if
$n\Omega = m\omega_r$, or (which is the same) if and only if the
 closure condition is
fulfilled,

$$ 
r = r_{\rm closure} = {3r_G \over {1 - n^2/m^2}},
\eqno (4)
$$
\noindent 
where $1 \le n < m$ are two integer numbers. 

\par
We see that the closure condition (4) leads to a similar (but
 different) quantization of
circular orbits as does the holonomy invariance condition (1). 
Rothman et al. noticed that
holonomy invariance occurs at circles outside of the free circular 
photon orbit, $r > r_{\rm ph} = (3/2)r_G$. 
They stressed that this could be intuitively understood in terms of the
optical geometry, i.e., 3-space conformally rescaled in such a way
 that light trajectories
are geodesic in it [4, 5]. The fact, most relevant in the context of
 holonomy on
timelike circular geodesics,
 that $r_{\rm ph}$ coincide with the circle of a minimal
circumference length, is an obvious triviality in optical geometry.
 Similarly, the closure
condition can only be fulfilled for circles outside the marginally
 stable orbit for test particles, $r > r_{\rm ms} = 3 r_G$. 
The innermost orbit for which the closure is possible
has a minimal curvature radius, a fact which again is most relevant,
 and trivial to understand in optical geometry.

\vskip 0.5truecm
\noindent
{\bf 3. Quantization of orbits in accretion disks: resonances and kHz QPOs}

\vskip 0.5truecm 
\noindent 
In numerous sources that astronomers directly observe, 
matter is being accreted onto the
central black black hole through a vertically thin accretion disk:
 the vertical thickness
$H$ is much smaller than the corresponding radius $r$.
 For inner parts of accretion disks
that are relevant in the present context, typically $h = H/r \approx
0.1$.
 In such disks,
and in the relevant range of radii (a few times $r_{ms}$),
 matter moves on orbits not much
different from circular geodesics: pressure, viscosity,
 radiation, magnetic field, and other
effects could be considered as being very small perturbations.
 Therefore, in thin accretion
disks, there {\it should be} present oscillation modes
 whose frequencies (in the considered range of radii)
 are very close to $\omega_r$, as well as other modes whose frequencies
are close to the meridional epicyclic frequency $\omega_{\theta}$, 
i.e., the frequency of
oscillatory motion of a test particle about the orbital plane. 
In Schwarzschild geometry,
$\omega_{\theta} = \Omega$, and in Kerr geometry
 $\omega_{\theta}$ is somewhat smaller than
$\Omega$. The functions $\Omega (r)$,
 $\omega_r (r)$, $\omega_{\theta} (r)$ are explicitly
known for both Schwarzschild and Kerr geometry (see e.g. [2, 6, 7]).

\par 
The general statement that the frequencies of global modes 
in thin accretion disks should be
determined by the space-time metric
 is well supported by detailed
studies of thin accretion disk oscillatory modes, or as it is called
 ``diskoseismology",
pioneered by Kato and developed by Wagoner and collaborators [2,6].
It has been suggested that prominent $g-$modes and $c$-modes can be
identified
with the ``stable'' frequencies observed in black hole candidates [8],
but the frequencies of these modes are not expected to be in rational
ratios. Other frequencies have also been investigated [2,6,7,9]. 
The well-known and important global $p$-modes have frequencies close
to $\omega_r$. Other, local, modes have frequencies
close to $\omega_{\theta}$ [10].
In these cases, pressure and other non-geodesic effects
contribute only to a small shift of the two frequencies.

\par
Our suggestion, is that the
two frequencies could be in a rational ratio,
 $n \omega_{\theta} = m \omega_r$, and when this
happens, a resonance may occur and the resulting oscillation is prominent
(and relatively easily detectable) for some, but not all,
values of $n/m$. Two of us 
have previously pointed out [11,12,13] that such
resonant frequencies
may indeed be observed as kilohertz quasi-periodic oscillations (QPOs)
 in X-ray emission of
accreting black holes and neutron stars (for a review of QPOs
properties see [14]), and show elsewhere [10]
that it follows naturally from this rational-ratio constraint and from 
fundamental properties of the relevant nonlinear oscillators
 that 3:2 and 5:3 ratios are expected among the observed frequencies,
as those connected
with the fastest growing modes. 
In the black hole source GRO J1655-40 the two high
frequencies observed are
450 Hz and 300 Hz [15], i.e., they are in a 3:2 ratio  [12,13], as expected
in our theory.  
The same 3:2 ratio of frequencies may have also been observed
in the black hole source XTE J1550-564 [16].
In the black hole source GRS 1915+105 the two frequencies observed
 are 70 Hz and 42 Hz [17],
i.e., they are in the also expected 5:3 ratio [12].

\par
We will not further discuss here our theory of excitation 
of such coherent
perturbations in accretion disks. 
We only point out that, most typically, the relative size
of the perturbation is of the order of $h$
 and therefore its relative surface of the order
$h^2 \approx 0.01$. If the resonant modes are excited,
 all properties of thin accretion
disks should oscillate, and in particular there should be
 an oscillatory part to
 the emitted X-ray flux,  of fractional amplitude $\approx h^2$
i.e., a few percent, as observed.
 The lifetime of the perturbation
is regulated not only by shear and dissipation in the disk,
 but also by the resonance
itself, and therefore it need not be of the order of the
 orbital period, but could be
much longer.

\par
 Note, that in the case of Schwarzschild geometry one 
has [2] $\omega_{\theta} =
\Omega$ and therefore the resonances occur at the ``quantized" 
orbits of radius given by Eq. (4). For the
Kerr black holes $\omega_{\theta} \neq \Omega$.  This is good news: 
the Kerr angular
momentum parameter $a$ appears directly in theory (governing some of
 its quantitative
aspects), and thus could be estimated from fitting theory to
observations [13, 18].

\par
At present, the QPO phenomenon is the only known strong-field consequence of
Einstein's relativity that could have been already directly observed 
(with an adequate precision),
and not mixed up with other effects that are difficult to follow 
in all relevant details. No
other observed phenomenon could rival QPOs in this respect. 
In particular, the oft-discussed  profiles of the Fe fluorescent line 
seen in X-ray spectra of some
active galactic nuclei, and believed to originate close to black holes
[19],
 in fact depend on many ``dirty astrophysics" factors that are difficult
 to estimate [20]. The
brilliant idea of Narayan, Lasota and others (see e.g. [21])
 that the observational fact
that accreting black holes sources are dimmer than accreting
 neutron stars [22] is a direct
evidence for the event horizon, suffers from the same 
difficulty---unknown factors
enter relevant details in a way impossible to control today [23].

\vskip 0.5truecm
\noindent
4. Conclusions

\vskip 0.5truecm
\noindent
Natural frequencies of oscillation of fluid on nearly circular orbits
 close to black holes
(and neutron stars) are a clean property of the strong gravity
 alone---as is
the closely related property of holonomy invariance. 
If Einstein was right, they simply
are there, and the theory of gravity need not  be supplemented
 by assumptions on the equation of
state, turbulence, magnetic field value, opacity, or any other
special conditions,
in order to explain the values of fundamental frequencies 
in oscillatory behaviour of matter
near black holes. There is a very serious possibility that 
the observed QPOs reflect
directly these frequencies. The QPO phenomenon is unique in 
its importance for testing
strong gravity.

\vskip 0.5truecm
\noindent
References:
\parindent=0pt

[~1] Rothman, T., Ellis, G.F.R., Murugan, J.,
Class. Quantum Grav. 18 (2001) 1217-1233

[~2] Kato, S., Fukue, J., Mineshige, S.,
{\it Black-hole accretion disks}, Kyoto University Press, Kyoto (1998)   

[~3] Amsterdamski, P., Bulik, T., Gondek-Rosi{\'n}ska, D., Klu{\'z}niak, W.,
Astr. Astrophys. 381 (2002) L21-L24

[~4] Abramowicz, M.A., Carter, B., Lasota J.-P.
Gen. Rel. Grav. 20 (1988) 1173-1183

[~5] Abramowicz, M.A.,
Mon. Not. Roy. astr. Soc. 256 (1992) 710-718

[~6] Wagoner, R.V., Physics Reports 311 (1999) 259

[~7] Kato, S.,
Pub. Astr. Soc. Jap. 52 (2000) 1125-1131

[~8] Wagoner, R. V., Silbergleit, A. S., Ortega-Rodr{\'\i}guez, M.,
Astrophys. J. 559 (2001) L25-L28

[~9]  Wallinder, F., Mon. Not. Royal Astron. Soc. 273 (1995) 1133-1140

[10] Abramowicz, M.A., Klu{\'z}niak, W., (2002) to be submitted to astro-ph

[11] Klu{\'z}niak W., Abramowicz, M.A., (2000) (available
 at http://xxx.lanl.gov/abs/astro-ph/0105057) manuscript
submitted to Phys. Rev. Lett. in December 2000,
before the observational results of
Strohmayer [15,16] were announced (April and June 2001), or indeed
submitted to ApJ Lett. (January and March 2001).

[12] Klu{\'z}niak W., Abramowicz, M.A., 
Acta Phys. Polonica B 32 (2001) 3605-3612

[13] Abramowicz, M.A., Klu{\'z}niak W.,
Astron. Astrophys.  374 (2001) L19-L20 

[14] van der Klis, M.,
Ann. Rev. Astron. Astrophys. 38 (2000) 717-760  
  
[15] Strohmayer, T.E.,
Astrophys. J. 552 (2001) L49-L53

[16] Miller, J.M., Wijnands, R., et al., Astrophys. J. 563 (2001) 928-933

[17] Strohmayer, T.E.,
Astrophys. J. 554 (2001) L169-L172. 

[18] Kato, S.,
Pub. Astr. Soc. Jap., 53 (2001) L37-L39

[19] Fabian, A. C.,
in Abramowicz M.A., Bj{\"o}rnsson, G., Pringle, 
{\it Theory of Black Hole Accretion Disks}, Cambridge University Press,
Cambridge (1998)

[20] Karas, V., Czerny, B., Abrassart, A., Abramowicz, M. A.,
Mon. Not. Roy. astr. Soc. 318 (2000) 547-560 

[21] Narayan, R., Mahadevan, R., Quataert, E.,
in Abramowicz M.A., Bj{\"o}rnsson, G., Pringle, 
{\it Theory of Black Hole Accretion Disks}, Cambridge University Press,
Cambridge (1998)

[22] Garcia, M.R., McClintock, J.E., Narayan, R., Callanan, P., Barret, D., Murray, S.S.,
Astrophys. J. 553 (2001) L47-L50

[23] Lasota, J.-P.,
invited talk at EDPS Conference, Lyon, 
Soc. Fran{\c c}aise d'Astronomie et d'Astrophysique,
Conference Series, 2002, p. 86 (and astro-ph/0110212)

\bye